\documentclass[12pt]{article}
\usepackage{epsfig}

\def\be{\begin{equation}}
\def\ee{\end{equation}}
\def\ba{\begin{array}{c}}
\def\ea{\end{array}}

\def\ben{$$}
\def\een{$$}

\newcommand{\kt}{\rangle}
\newcommand{\br}{\langle}
\begin{document}


\vspace{.35cm}

 \begin{center}{\Large \bf

Scattering theory with localized non-Hermiticities

  }\end{center}

\vspace{10mm}

 \begin{center}

 {\bf Miloslav Znojil}

 \vspace{3mm}
Nuclear Physics Institute ASCR,

250 68 \v{R}e\v{z}, Czech Republic

{e-mail: znojil@ujf.cas.cz}

\vspace{3mm}


\end{center}

\vspace{5mm}


\section*{Abstract}

In the context of the recent interest in solvable models of
scattering mediated by non-Hermitian Hamiltonians (cf. H. F.
Jones, Phys. Rev. D 76, 125003 (2007)) we show that the well known
variability of the {\it ad hoc} choice of the metric $\Theta$
which defines the physical Hilbert space of states can help us to
clarify several apparent paradoxes. We argue that with a suitable
$\Theta$ a fully plausible physical picture of the scattering can
be recovered. Quantitatively, our new recipe is illustrated on an
exactly solvable toy model.

\newpage

\section{\label{Ia}Introduction}

Whenever one considers the one-dimensional differential
Schr\"{o}dinger equation
 \be
 \left [
 -\frac{d^2}{dx^2}+V(x) \right ]\,\psi(x)=E\,\psi(x)\,,
 \ \ \ \ \ \ \ x \in (-\infty,\infty)
 \label{SE}
 \ee
in the scattering regime, i.e., with the boundary conditions
describing the transmitted and reflected waves,
 \be
 \psi(x) =
 \left \{
 \begin{array}{lr}
 e^{i\kappa x}+C\,e^{-i\kappa x}\,,\ \ \ \ \
 & x \ll -1\,,\\
 D\,e^{i\kappa x}\,,\ \ \ \ \ \ \ \ \ \ \
 &x \gg 1\,
 \ea
 \right .
 \label{scatbc}
 \ee
one usually assumes that the flow of probability is conserved,
$|C|^2+|D|^2=1$. Recently, Jones \cite{Jones} pointed out that
several serious conceptual difficulties can arise when certain
current tacit assumptions (demanding, typically, the reality of
the potential $V(x)$) are, tentatively, weakened. He also
presented several persuasive arguments {\it why} one should try to
weaken these assumptions.

Among the latter arguments, the most persuasive support of
theoretical as well as conceptual innovations would certainly lie
in the undeniable success of the recent, phenomenologically
motivated transition to certain manifestly non-Hermitian
Hamiltonians generating real and observable spectra of bound
states~\cite{BB}. The idea (well known to mathematicians
\cite{BG}) found, recently, several interesting applications in
nuclear physics \cite{Geyer} and in field theory \cite{Carl}.

Of course, there exists an obvious difference between the
bound-state problem (for which wave functions $\psi(x)$ are
localized) and the scattering scenario (where all the waves remain
non-negligible all along the whole real axis). Jones  \cite{Jones}
even came to a rather sceptical  conclusion that the preservation
of a sensible probabilistic interpretation of a generic
non-Hermitian model of scattering may be quite costly and
difficult even when the tentative introduction of a suitable
non-Hermiticity in the Hamiltonian itself remains restricted to a
very small domain of $x$. Similar observations have been also made
in a few older scattering models where the violation of the rule
$|C|^2+|D|^2=1$ has been explained and interpreted as a
phenomenologically acceptable manifestation of the presence of
some ``hidden" degrees of freedom in the model \cite{Ahmed}.

Being not satisfied by these ``effective" theories the author of
ref.~\cite{Jones} formulated a much more ambitious project where
the physical picture of scattering  would parallel the
above-mentioned ``fundamental" theory of bound states based on
non-Hermitian Hamiltonians \cite{Geyer,Carl}. Unfortunately, the
quantitative results of ref.~\cite{Jones} were not too encouraging
(cf. also their recent completion \cite{Jonesdva}). In essence,
the sensible probabilistic interpretation of the models under
consideration seemed to require that the above-mentioned standard
boundary conditions should be modified to read
 \be
 \psi(x) =
 \left \{
 \begin{array}{lr}
 e^{i\kappa x}+C\,e^{-i\kappa x}\,,\ \ \ \ \
 & x \ll -1\,,\\
 D\,e^{i\kappa x}+D'\,e^{-i\kappa x}\,,\ \ \ \ \ \ \ \ \ \ \
 &x \gg 1\,.
 \ea
 \right .
 \label{sctbc}
 \ee
Unfortunately, this formula contains a strongly counterintuitive
``backwards-running" component proportional to $D' \neq 0$ in the
scattered solution.

In what follows we intend to weaken  the resulting skepticism. We
shall start from the same basic theoretical premise and postulate
that the effective theories of refs. \cite{Ahmed} are in fact not
too interesting since they just mimic the presence of certain
unknown dynamical mechanisms admitting, i.a., an annihilation
and/or creation during the scattering. In this sense, we intend to
search now for a new quantitative support for the possible {\em
feasibility and consistency} of the alternative {\em fundamental}
approach where one tries to re-establish the conservation of the
probability.

In section \ref{Jonesovo} we shall commence our considerations by
a brief and compact review of the most relevant results of
ref.~\cite{Jones}. We summarize there the overall philosophy of
the fundamental theory where the input Hamiltonian $H$ [say, of
eq.~(\ref{SE})] is interpreted as a mere {\em auxiliary} operator.
One assumes that with this {\em auxiliary} operator the {\em
formal} calculations become {\em exceptionally} simple. At the
same time, the ``correct" physics is assumed to be defined, via an
invertible map $\Omega$, in terms of a certain ``true" physical
Hamiltonian
 \be
 h = \Omega\,H\,\Omega^{-1}\,
 \label{haha}
 \ee
which is expected to be {\em prohibitively complicated}. For
illustration one can recollect the ``true nuclear physics" of
ref.~\cite{Geyer} where $h$ was a full-fledged Hamiltonian of an
atomic nucleus while $H$ has been constructed as its much more
easily tractable map.

In all the similar scenarios, the similarity mapping $\Omega$ must
be assumed {\em non-unitary} -- otherwise, one would return to the
mere traditional, Hermitian class of Hamiltonians. In order to
suppress or circumvent the related, mostly purely technical
obstacles we decided to employ our recent bound-state experience
\cite{Jednap} and to restrict our attention to a fairly restricted
class of the solvable illustrative dynamical models. They are
introduced and described in section \ref{RK2}.

The manifestly non-perturbative character of our present class of
models will enable us to make use of the flexibility of the
mappings $\Omega$ and to draw several consequences from the exact
solvability of our models. In section \ref{RK4} we shall show how
some of the apparently unavoidable paradoxes of ref. \cite{Jones}
can find their explanation and resolution. In particular, for our
manifestly non-Hermitian set of specific Hamiltonian-representing
operators $H\neq H^\dagger$ we shall demonstrate that they can be
assigned a consistent and unitary physical interpretation of the
scattering based on standard asymptotic boundary conditions
(\ref{scatbc}).

In section \ref{summ} a compact summary of our present
constructive arguments will be complemented by a few optimistic
remarks concerning the possible future extension of the class of
our present models of scattering towards some less schematic
non-Hermitian Hamiltonians.

\section{\label{Jonesovo}Scattering from complex potentials }

The mathematical background of the fundamental models of
scattering from localized non-Hermitian centers will be
illustrated here on a set of solvable models. On this level we
shall demonstrate that a very natural interpretation of this type
of scattering is feasible. On an abstract phenomenological level
we shall stress that in our present update of the extension of the
scattering theory of ref.~\cite{Jones} a core of observational
consistency {\em should and can} be sought in an at least partial,
asymptotic survival of the observability of the coordinates.

\subsection{Jones' solvable example}

For the majority of the real and smooth one-dimensional
short-range potentials the description of the scattering is
routine. One solves the ordinary differential Schr\"{o}dinger
equation under the standard asymptotic boundary conditions. For
the complex $V(x)$s in (\ref{SE}), {fundamental theory} makes the
scattering unitary via an appropriate adaptation of the metric
$\Theta$ in the Hilbert space of states. Unfortunately, an
unexpected and unpleasant consequence has been detected in
\cite{Jones} where the replacement of the standard asymptotic
boundary conditions (\ref{scatbc}) by their fairly
counterintuitive ``amendment" (\ref{sctbc}) has been found
necessary in principle. Fortunately, in a concrete illustration
using the complex delta function toy potential of
ref.~\cite{delta},
 \be
 V^{(Mostafazadeh)}(x) = 2\lambda
 (1+{\rm i}\varepsilon)\,\delta(x)
 \label{aliho}
 \ee
the contribution of the nonvanishing coefficient $D'$ to the flow
of probabilities proved negligible \cite{Jones}. Thus, in the
leading-order approximation it was possible to return to the
original boundary conditions (\ref{scatbc}).

After such an approximate confirmation of the internal consistency
of the fundamental theory approach a careful perturbation analysis
of scattering by potential (\ref{aliho}) has been performed in
\cite{Jones} leading to the intermediate result
  \be
 |C|^2+|D|^2=
 \left (
 1-\frac{2\varepsilon\,q}{1+\varepsilon^2+q^2}
 \right )^{-1}\,, \ \ \ \ \ q=\frac{\sqrt{E}}{\lambda}\,.
 \label{jineseho}
 \ee
In the naive effective-theory interpretation of this formula the
non-conservation of the flow of probability  merely reflects the
fact that the  manifestly non-Hermitian Hamiltonian $H$ is merely
an auxiliary operator defined in the ``wrong" Hilbert space ${\cal
H}^{(unphysical)}$. One has to employ the map $\Omega$ to move to
another, {\em unitarily non-equivalent} ``correct" Hilbert space
${\cal H}^{(physical)}$ where the ``true" representant
(\ref{haha}) of the Hamiltonian remains safely Hermitian
\cite{Jones}.

\subsection{ \label{RK}Long-range non-localities induced by the
short-range potential: a paradox}

One of the most unpleasant formal features of physical metric
$\Theta=\Theta(H)$ \cite{footnotea} in the Hilbert space of states
${\cal H}^{(physical)}$ is that it is usually strongly non-local
even if the original potential is local, $V=V(x)$ (cf. also
\cite{Jednap}). This implies that it is hardly feasible to perform
any computations in ${\cal H}^{(physical)}$. In the purely
auxiliary Hilbert space ${\cal H}^{(unphysical)}$ the computations
are assumed much easier. All the ket vectors $|\psi\kt$ which lie
in the latter space lack, unfortunately, any direct physical
interpretation. Even the standard requirement of the observability
of the Hamiltonian degenerates, in this auxiliary space, to the
identity \cite{Geyer}
 \be
 H^\dagger = \Theta\,H\,\Theta^{-1}\,.
 \label{tarov}
 \ee
This relation represents the Hermiticity of the Hamiltonian $h$ in
${\cal H}^{(physical)}$,
 \ben
 h=\Omega\,H\,\Omega^{-1}=h^\dagger=
\left (\Omega^{-1}\right )^\dagger\,H^\dagger\,\Omega^\dagger
 \,.
 \een
We may deduce that one has to put $\Theta \equiv \Omega^\dagger
\Omega$ \cite{SIGMA}. In the language of mathematics we must
guarantee that the physical metric  $\Theta$ is compatible with
relations (\ref{tarov}). {\it Vice versa}, any operator
$\sigma=\sigma^\dagger$ representing an observable in ${\cal
H}^{(physical)}$ has to have its appropriate quasi-Hermitian
partner $\Sigma$ in ${\cal H}^{(unphysical)}$ which obeys an
analogue of eq.~(\ref{tarov}) using the {\em same}
metric~\cite{cc}.

The shift $\varepsilon$ in eq.~(\ref{aliho}) has been assumed
small in ref.~\cite{Jones}. This opened the possibility of an
explicit use of Mostafazadeh's metric \cite{delta} available in
perturbation-series form
 \be
 \Theta^{(Mostafazadeh)} \,\equiv\,\eta
 =I +\varepsilon\, \eta^{(1)} + {\cal O}(\varepsilon^2)\,.
 \label{jednicka}
 \ee
In coordinate representation the unit operator $I$ becomes
represented by the delta-function kernel $\delta(x-y)$ but a {\em
manifest and large nonlocality} emerges already in the first
perturbation order,
 \be
 \eta^{(1)}(x,y)=\frac{\lambda}{2}\,{\rm sgn}\,(y^2-x^2)\,
 \left [\theta(xy)\,e^{-\lambda\,|x-y|}+\theta(-xy)\,e^{-\lambda\,|x+y|}
 \right ]\,.
 \label{takyjednicka}
 \ee
The emergence of such a nonlocality leads  to serious problems
because ``the physical picture of the scattering is completely
changed" since, on the positive half-axis the physical wave
function ``no longer represents a pure outgoing wave \ldots but
\ldots contains an ${\cal O}(\varepsilon)$ component of an
incoming wave as well" \cite{Jones}.

We shall be able to show that while the deformations caused by the
use of the {\em locally} non-Hermitian interaction remain
long-ranged in their character, they need not necessarily lead to
the emergence of  spurious components in the outcoming wave. Such
an observation is not in contradiction with the fact that ``one
should change the Hilbert space by adopting the appropriate metric
[which] must differ from the standard one not only in the vicinity
of the non-Hermitian potentials, but also at distances remote from
it" \cite{Jones}. Nevertheless, we shall argue that at least some
of the spuriosities emerge {\em only} due to an inappropriate
choice of a specific metric, the  definition of which is known to
contain infinitely many free parameters \cite{Geyer,SIGMA,Ali}. In
this sense we shall make use of a simpler model and recommend here
the construction and use of {\em another}, quasi-local metric
operator $\Theta=\Theta^{(QL)} \neq \eta$.

\section{\label{RK2} Discrete Schr\"{o}dinger equations}

Some of the standard {scattering-theory} considerations can be
simplified when one replaces the ordinary differential equation
(\ref{SE}) by the difference equation
 \be
 -\frac{\psi(x_{k-1})-2\,\psi(x_k)+\psi(x_{k+1})}{h^2}+V(x_k)\,
 \psi(x_k)
 =E\,\psi(x_k)\,.
 \label{SEdis}
 \ee
For example, in some pragmatic numerical calculations one chooses
a sufficiently small step-size $h> 0$ and introduces  discrete
coordinates,
 \be
  x_k=k\,h\,,\ \ \ \ \
 \ \ \ \ \ k = 0, \pm 1, \ldots\,.
 \label{kkoo}
 \ee
This makes the usual real line replaced or  approximated by an
infinitely long discrete lattice. The most elementary application
of such a discretization  occurs when one wants to construct bound
states. For certain real as well as complex potentials a sample of
the construction may be found in our papers~\cite{I}. Some of them
illustrate also the fundamental-theory approach to the
non-Hermitian quantum bound states where the Hamiltonian $H$ is
treated as quasi-Hermitian, i.e., Hermitian only in the
Hamiltonian-dependent Hilbert space ${\cal H}^{(physical)}$.

\subsection{Discrete in and out free waves }

Let us assume that the potential in eq.~(\ref{SEdis}) vanishes
beyond certain not too large distance from the origin, $ V(x_{\pm
j})=0$, $j = M, M+1, \ldots $. In the free-motion domain,  we
abbreviate $\psi_j=\psi(x_j)$ and $2\,\cos \varphi=2-h^2 E$  and
replace eq.~(\ref{SEdis}) by recurrences
 \be
 -\psi_{j-1}^{(0)}+2\,\cos \varphi\,\psi_j^{(0)}
 -\psi_{j+1}^{(0)}=0\,
 \label{SEdisc}
 \ee
or by the matrix equations
$H_0\,\vec{\psi}^{(0)}=h^2k^2\,\vec{\psi}^{(0)}$ or
$M_0(\varphi)\,\vec{\psi}^{(0)}= 0$, viz.,
 \be
 \left (
 \begin{array}{ccccc}
 \ddots&\ddots&\ddots &\vdots&\\
 \ddots&2\cos \varphi &-1& 0&\ldots \\
 \ddots&-1&2\cos \varphi&-1&\ddots\\
  \ldots&0&-1&2\cos \varphi&\ddots\\
 & \vdots& \ddots&\ddots&\ddots
 \ea
 \right )
 \,\left (
 \ba
 \vdots\\
 \psi_{j-1}^{(0)}\\
 \psi_j^{(0)}\\
 \psi_{j+1}^{(0)}\\
 \vdots
 \ea
 \right )=0\,
 \label{bigmat}
 \ee
which may be assigned a doublet of independent solutions,
 \ben
 \psi_k^\pm =const\,\cdot\,\varrho^k_\pm\,,\ \ \ \ \ \varrho_{\pm}=
 \exp(\pm i\,\varphi)\,.
 \een
Precisely in the spirit of ref.~\cite{Carl} we can speak about a
${\cal PT}-$symmetric free Hamiltonian,
 \ben
 H_0=
 \left [\begin {array}{rrrrr}
  \ddots&\ddots&&&\\{}
  \ddots&2&-1&&
 \\{}
 &-1&2&-1&\\{}&&-1&2&\ddots
 \\{}
 &&&\ddots&\ddots\end {array}\right ]\,,
 \ \ \ \ \
 {\cal P}=
 \left [\begin {array}{rlcrr}
  &&&&\ \ \dot{ \dot{\dot{}\   }\  }\\{}
  &&&\  1 &
 \\{}
 && 1 &&\\{}&1 \ &&&
 \\{}\ \ \ \
\dot{ \dot{\dot{}\   } \ }
 &&&&\end {array}\right ]\,.
 \een
It is easy to verify that  its spectrum is real, provided only
that the new energy variable $\varphi$ is real \cite{Jednap}. This
imposes an inessential constraint $-2\leq 2-h^2 E \leq 2$ upon the
energy range, i.e., we must have $E \in (0, 4/h^2)$. At any finite
choice of the lattice step $h>0$ this is reminiscent of the
similar feature of the spectra in relativistic quantum systems.
This connection has been given a more quantitative interpretation
in ref.~\cite{jaotom}.

Let us finally add an interaction with nonzero elements forming
merely a finite-dimensional submatrix in the Hamiltonian. The
scattering of an incoming wave may be then characterized, say, by
the boundary conditions
 \be
 \psi(x_m) =
 \left \{
 \begin{array}{ll}
 e^{i\,m\,\varphi}+R\,e^{-i\,m\,\varphi}\,,\ \ \ \ \
 \ \ \ \ \ \
 & m \leq -M\,,\\
 T\,e^{i\,m\,\varphi}\,,\ \ \ \ \ &
 m \geq M-1\,
 \ea
 \right .
 \label{discatbc}
 \ee
i.e., by eq.~(\ref{scatbc}) in its discrete version.

\subsection{Discrete short-range model of
scattering\label{RK11}}

In a way inspired by our recent studies of finite-dimensional
non-Hermitian Hamiltonian matrices $H$ with real spectra
\cite{Jednap,dva} let us pick up one of these models and
contemplate its infinite-dimensional generalization which would
admit scattering solutions. Its explicit matrix representation
will be tridiagonal, one-parametric and doubly infinite,
 \be
 H_1=
 \left [\begin {array}{rrrc|crrr}
  \ddots&\ddots&&&&&&
  \\{}
  \ddots&2&-1&&&&&
 \\{}
 &-1&2&-1&&&&
 \\{}
 &&-1&2&-1-a&&&
 \\
  \hline
 &&&  -1+a&2&-1&&
 \\{}
 &&&&-1&2&-1&
 \\{}
 &&&&&-1&2&\ddots
 \\{}
 &&& &&&\ddots&\ddots
 \end {array}\right ]\,.
 \label{tridif}
 \ee
Inside the interval of $a \in (-1,1)$ all the $2K-$dimensional
truncations of $H_1$ have the $2K-$plets of eigenvalues which are
all real and lie inside the closed interval $(1,3)$.  In the
non-Hermitian regime with the growing $|a|$ these energies
pairwise degenerate at the ``exceptional points" $a=\pm 1$ and get
complex at $|a|>1$. At $K=2$ and $K=3$ the smoothness of the
$a-$dependence of these truncation-dependent standing wave
energies has been illustrated in \cite{Jednap}. At $K=1$ their
explicit form reads $h^2k^2_\pm = 3\pm \sqrt{1-a^2}$ and offers a
schematic guidance and a nice quantitative illustration of what
happens in general.

In a search for the transmission and reflection amplitudes our
infinite-dimensional matrix problem for scattering
$H_1\,\vec{\psi}^{(1)}=h^2k^2\,\vec{\psi}^{(1)}$ can be split in
its two free-motion parts~(\ref{bigmat}) for the respective ``in"
and ``out" solutions (\ref{discatbc}) valid up to $M=1$. They have
to be matched near the origin,
 \ben
  \left [\begin {array}{rccr}
  -1&e^{{\rm i}\varphi}+ e^{-{\rm i}\varphi} &-1-a&0
 \\
 0&  -1+a&e^{{\rm i}\varphi}+ e^{-{\rm i}\varphi}&-1
 \end {array}\right ]\,
  \left [\begin {array}{c}
 R\,e^{2{\rm i}\varphi} \\
 R\,e^{{\rm i}\varphi} \\
  T \\
  T\,e^{{\rm i}\varphi } \\
 \end {array}\right ]=
  \left [\begin {array}{rc}
  1&-e^{{\rm i}\varphi}- e^{-{\rm i}\varphi}
 \\
 0&  1-a
 \end {array}\right ]\,
  \left [\begin {array}{c}
  e^{-2{\rm i}\varphi} \\
  e^{-{\rm i}\varphi}
 \end {array}\right ]
 \,.
 \een
These two linear equations for $R$ and $T$ can be simplified,
 \ben
  \left [\begin {array}{c}
 1\\
 -(1-a)e^{{\rm i}\varphi}
 \ea
 \right ]\,
 R+
  \left [\begin {array}{c}
 -(1+a)\\
 e^{-{\rm i}\varphi}
 \ea
 \right ]\,
 T=
  \left [\begin {array}{c}
 -1\\
 (1-a)e^{-{\rm i}\varphi}
 \ea
 \right ]\,.
 \een
It is easy to write down their explicit solution,
 \ben
 R=-\frac{a^2}{\triangle}\,,
 \ \ \ \ \
 T=\frac{(1-a)(1- e^{2{\rm i}\varphi})}{\triangle}\,,
 \ \ \ \
 \triangle = 1-(1-a^2)\, e^{2{\rm i}\varphi}\,.
 \een
This gives an exact analogue
 \be
 |R|^2+|T|^2=\frac{1-{a\,}{[1+
 U(a,\varphi)]^{-1}}}{1+{a\,}{[1+U(a,\varphi)]^{-1}}}
 \,,\ \ \ \ \ \
  U(a,\varphi)=\frac{a^4}{2\,(1-a)\,(1-\cos 2  \varphi)}\,
  \label{uvel}
 \ee
of equation Nr. (11) of ref.~\cite{Jones}. In both cases, the sum
of probabilities is greater than 1 or less than 1 depending on the
sign of the deviation of the coupling constant form its Hermitian
zero limit. The same conclusion can be read in ref.~\cite{Jones}
so that in the weak-coupling regime our present
difference-operator parameter $a$ plays the same dynamical role as
its differential-operator predecessor $\varepsilon$ in
eq.~(\ref{aliho}). Moreover, due to the nonperturbative character
of our result one can rewrite eq.~(\ref{uvel}) in the equivalent
form \cite{Jonespr}
\begin{equation}
|R^2|+|T|^2=\frac{a^4+4(1-a)^2\sin^2{\varphi}}{a^4+4(1-a^2)\sin^2{\varphi}}
\end{equation}
which is more compact and clarifies the nature of the singularity
reached in the limit $a \to 0$.

We shall also see below (cf. section \ref{existence}) that after
the necessary adaptation of the Hilbert space of states and after
the {\em ad hoc} modification of the inner product the net result
of the changes will be the elementary modification of the
coefficient of $\sin^2{\varphi}$ in the numerator to $4(1-a^2)$,
thus restoring the usual and physically consistent unitarity of
the scattering.

\subsection{More-parametric models \label{RK3} }


With another Hamiltonian
 \ben
 H_2=
 \left [\begin {array}{rrc|cc|crr}
  \ddots&\ddots&&&&&&
  \\{}
  \ddots&2&-1&&&&&
 \\
 &-1&2&-1-b&&&&
 \\
 \hline
 &&-1+b&2&-1-a&&&
  \\{}
 &&&  -1+a&2&-1-b&&
 \\
 \hline
 &&&&-1+b&2&-1&
 \\
 &&&&&-1&2&\ddots
 \\{}
 &&& &&&\ddots&\ddots
 \end {array}\right ]
 \een
one can expect that many results obtained previously for its $b=0$
special case $H_1$ can find a natural generalization. The same
expectations concern also the next-step candidate with three free
parameters,
 \ben
 H_{3}=
 \left [\begin {array}{rccc|cc|cccr}
\ddots & \ddots&\ddots&&&&&&&
  \\{}
  &-1&2&-1-c&&&&&&
 \\
 &&-1+c&2&-1-b&&&&&
 \\
 \hline
 &&&-1+b&2&-1-a&&&&
  \\{}
 &&&&  -1+a&2&-1-b&&&
 \\
 \hline
 &&&&&-1+b&2&-1-c&&
 \\
 &&&&&&-1+c&2&-1&
 \\{}
 &&& &&&&\ddots&\ddots&\ddots
 \end {array}\right ]
 \een
etc. In all of them the asymptotic ``in" and ``out" solutions of
eq.~(\ref{discatbc}) remain uninfluenced by the interaction.
Equally well, the matching of these in and out solutions remains
feasible at any number $k$ of parameters.

In our first nontrivial scattering model
$H_2\,\vec{\psi}^{(2)}=h^2k^2\,\vec{\psi}^{(2)}$ the matching may
be mediated by the choice of $M=2$. This means that the following
four matching conditions must be considered,
 \ben
\left [\begin {array}{rccccr}
    -1&2\cos \varphi&-1-b&0&0&0
 \\
 0&-1+b&2\cos \varphi&-1-a&0&0
  \\{}
 0&0&  -1+a&2\cos \varphi&-1-b&0
 \\
 0&0&0&-1+b&2\cos \varphi&-1
 \end {array}\right ]\,
  \left [\begin {array}{c}
   e^{-3{\rm i}\varphi}+
 R\,e^{3{\rm i}\varphi} \\ e^{-2{\rm i}\varphi}+
 R\,e^{2{\rm i}\varphi} \\ e^{-{\rm i}\varphi}+
 R\,e^{{\rm i}\varphi}-
 \chi_{-1}
 \\
 T+\chi_0
 \\
  T\,e^{{\rm i}\varphi } \\
  T\,e^{2{\rm i}\varphi }
 \end {array}\right ]=0
 \,.
 \een
The first line defines the correction $\chi_{-1}$ and the last
line defines the correction $\chi_{0}$,
 \ben
 (1+b)\,\chi_{-1}=b\,(e^{-{\rm i}\varphi}
 +R\,e^{{\rm i}\varphi})\,,\ \
 \ \ \ \ \
 (1-b)\,\chi_{0}=b\,T\,.
 \een
This reduces the number of our equations to two again,
 \ben
\left [\begin {array}{cccc}
 -1+b&2\cos \varphi&-1-a&0
  \\{}
 0&  -1+a&2\cos \varphi&-1-b
 \end {array}\right ]\,
  \left [\begin {array}{c}
   e^{-2{\rm i}\varphi}+
 R\,e^{2{\rm i}\varphi} \\
 \left (
 e^{-{\rm i}\varphi}+
 R\,e^{{\rm i}\varphi}
 \right )/
 (1+b)
  \\
 T/
 (1-b)
 \\
  T\,e^{{\rm i}\varphi }
 \end {array}\right ]=0
 \,.
 \een
After their simplification we may easily eliminate
 \ben
 \frac{(1+b)\,T}{(1-a)\,(1-b)}=
 \frac{1+R\,e^{2{\rm i}\varphi}}{1+b^2\,e^{2{\rm i}\varphi}}\,.
 \een
We end up quickly with the explicit definition of $R$ for our
second model $H_2$,
 \ben
 R=-\frac{a^2+2\,b^2\cos 2\,\varphi + b^4}{\triangle}\,,
 \ \ \ \ \ \triangle= 1-(1-a^2)\, e^{2{\rm i}\varphi}
 +2b^2\,e^{2{\rm i}\varphi}
  +b^4\,e^{4{\rm i}\varphi}
 \,.
  \een
We observe a close parallelism with the preceding model. From the
easy first-order estimates
  \ben
  R ={\cal O}(a^2)+{\cal O}(b^2)\,,\ \ \ \ \ \
 \frac{(1+b)\,T}{(1-a)\,(1-b)}=
 1
 +{\cal O}(a^2)+{\cal O}(b^2)
 \,
 \een
we may immediately deduce that
 \be
 |R|^2+|T|^2=
 1-2\,a-4\,b
 +{\cal O}(a^2)+{\cal O}(b^2)
 \label{druhama}
 \,.
 \ee
This two-parametric dependence parallels closely the
one-parametric prediction offered by eq.~(\ref{uvel}).

\section{Unitarily non-equivalent Hilbert spaces \label{RK4} }

In a climax of our paper we shall make use of the fact that our
models are exactly solvable, at least in terms of the methods
based on the computer-assisted symbolic manipulations and
extrapolations. This will enable us to construct {\em many }
eligible candidates $\Theta$ for the metric in the physical
Hilbert space. In contrast, even the construction of their
subclass denoted by the symbol $\rho$ and possessing an explicit
perturbation form
 \be
 \varrho=I +\frac{1}{2}\,\varepsilon\, \eta^{(1)}
 + {\cal O}(\varepsilon^2)\,.
 \label{jednericka}
 \ee
was an achievement for differential operators in refs.
\cite{Jones} and \cite{delta}. In such a context, a core of our
present message is that due to the simplified, difference-operator
representation of observables we shall be able to select a better
metric $\Theta^{(QL)} \neq \eta$. With its use, some of the most
counterintuitive manifestations of the nonlocality paradox will
simply disappear.

\subsection{Ambiguity problem }

In the  effective interaction scenario, formulae (\ref{jineseho})
or (\ref{uvel}) and (\ref{druhama}) would certainly indicate the
presence of an absorption and/or creation at $\varepsilon \neq 0$
or $a\neq 0$ and $b \neq 0$. In the fundamental theory one assumes
a change of the Hilbert space such that the original (i.e.,
standard) definition of the inner product
 \be
 \br \psi|\psi'\kt = \int_{I\!\!R}\,\psi^*(x)\,\psi'(x)\,dx
 \label{usuall}
 \ee
is replaced by its more general weighted version in the new space,
 \be
 \br \psi|\psi'\kt_\Theta = \int_{I\!\!R^2}\,\psi^*(x)\,
 \Theta(x,x')\,\psi'(x')\,dx\,dx'\,.
 \label{newh}
 \ee
The purpose of such a change is in making the Hamiltonian
selfadjoint.

It is well known that the choice of the inner product (\ref{newh})
is ambiguous \cite{Geyer}. One of the standard constructive
solutions of the ambiguity problem giving $\Theta=\eta$ has been
proposed by Mostafazadeh \cite{delta}. In the mathematically most
easily tractable dynamical regime of a very small deviation
$|\varepsilon| \ll 1$ from Hermiticity this author arrived at the
explicit perturbation approximation (\ref{jednicka}) +
(\ref{takyjednicka}) where the maxima of function $\eta(x,y)$ lie
on the two perpendicular lines defined by the trivial equations
$x\pm y=0$ in the $x-y$ plane. Subsequently, the latter recipe has
been used in ref.~\cite{Jones} where operator
$\Omega^{(Jones)}\,\equiv\,\varrho$ was defined as a self-adjoint
square root (\ref{jednericka}) of  metric $\eta$ emphasizing that
in terms of physics, ``the relevant wave function is not $\psi(x)
\,\equiv\,\br x|\psi\kt$, but $\Psi(x) \,\equiv\,\br x|\Psi\kt =
\br x|\varrho|\psi\kt$".

In this context, the mathematical essence of our present amendment
of scattering theory lies precisely in an innovation of the choice
of $\Omega$ and $\Theta$ since among all the available mappings
$\Omega$ the selected $\varrho$ remains also very strongly
non-local, indeed.

\subsection{The existence of {\em diagonal} matrices $\Theta=\Theta^{(QL)}$
\label{existence}}

In the technically most complicated part of our present
considerations we decided to choose a Hamiltonian and to treat its
quasi-Hermiticity condition (\ref{tarov}) as a linear set of
equations for all the matrix elements of the metric.

In the first attempt we choose $H_1$ and verified that there
exists the infinite-dimensional matrix solution $\Theta_1$ of
eq.~(\ref{tarov}) which is {\em diagonal}, i.e., in our present
terminology, quasi-local,
 \ben
 \Theta_1^{(QL)}=
 \left [\begin {array}{rrrc|crrr}
  \ddots&&&&&&&
  \\{}
  &1-a&&&&&&
 \\{}
 &&1-a&&&&&
 \\{}
 &&&1-a&&&&
 \\
  \hline
 &&&&1+a&&&
 \\{}
 &&&&&1+a&&
 \\{}
 &&&&&&1+a&
 \\{}
 &&& &&&&\ddots
 \end {array}\right ]\,.
 \een
This result was obtained via tedious symbolic manipulations on the
computer. Its simplicity is both very surprising and very useful
because one of the integrations in the related inner product
(\ref{newh}) drops out. Moreover, its diagonal kernel can
trivially be factorized into the product of two diagonal operators
$\rho=\sqrt{\Theta}$, i.e.,
 \ben
 \rho_1^{(QL)}=
 \left [\begin {array}{rrrc|crrr}
  \ddots&&&&&&&
  \\{}
  &\sqrt{1-a}&&&&&&
 \\{}
 &&\sqrt{1-a}&&&&&
 \\{}
 &&&\sqrt{1-a}&&&&
 \\
  \hline
 &&&&\sqrt{1+a}&&&
 \\{}
 &&&&&\sqrt{1+a}&&
 \\{}
 &&&&&&\sqrt{1+a}&
 \\{}
 &&& &&&&\ddots
 \end {array}\right ]\,.
 \een
They remain self-adjoint and positive definite at all the not too
large real $a$s.

The diagonality of the latter matrix enables us to insert it in
eq. Nr. (17) of ref.~\cite{Jones} and to deduce that the explicit
formula for the ``correct" operator $X$ of the observable
coordinate {\em coincides} with its standard diagonal-matrix form
with elements given by eq.~(\ref{kkoo}) above. In the same manner
one can also recall eq.~(\ref{haha}) and introduce the operator
$h_1^{(QL)} = \rho_1^{(QL)}\,H_1\,\left (\rho_1^{(QL)}\right
)^{-1}$ which represents the isospectral {\em Hermitian}
Hamiltonian of our system and which replaces eq.~(\ref{tridif}) by
the real and symmetric tridiagonal matrix
 \be
 h_1^{(QL)} =
 \left [\begin {array}{rrrc|crrr}
  \ddots&\ddots&&&&&&
  \\{}
  \ddots&2&-1&&&&&
 \\{}
 &-1&2&-1&&&&
 \\{}
 &&-1&2&-\sqrt{1-a^2}&&&
 \\
  \hline
 &&&  -\sqrt{1-a2}&2&-1&&
 \\{}
 &&&&-1&2&-1&
 \\{}
 &&&&&-1&2&\ddots
 \\{}
 &&& &&&\ddots&\ddots
 \end {array}\right ]\,.
 \label{tridifve}
 \ee
We see that this operator differs from the purely kinetic
Hamiltonian just in a small vicinity of the scattering center. In
spite of such a strict locality of the interaction, the metric
itself remains deformed far away from the scattering center.

The fact that the manifest non-Hermiticity of our toy model $H_1$
did not involve the mixing of incoming and outgoing waves that
occurred in the model of ref.~\cite{Jones} encouraged us to
proceed towards the more complicated models using the same
brute-force method. After the choice of the next, two-parametric
non-Hermitian Hamiltonian $H_2$ the calculations still remained
sufficiently easy for us to deduce and verify the existence of the
following two-parametric quasi-local solution $\Theta_2^{(QL)}$ of
eq.~(\ref{tarov}) represented by the diagonal matrix
 \ben
 \left [\begin {array}{rrrc|crrr}
  \ddots&&&&&&&
  \\{}
  &\multicolumn{2}{c}{(1-a)\,(1-b)^2}&&&&&
 \\{}
 &&\multicolumn{2}{c|}{(1-a)\,(1-b)^2}&&&&
 \\
  \hline
 &&&(1-a)\,(1-b^2)&&&&
 \\%
 &&&&\multicolumn{2}{c}{ (1+a)\,(1-b^2)\ \ \ \ }
 &&
 \\
  \hline
 &&&& &\multicolumn{2}{c}{\ \ \ \ \ \ \ \ \ \  (1+a)\,(1+b)^2 }&
 \\{}
 &&&&&&\multicolumn{2}{c}{(1+a)\,(1+b)^2\ \ \ \ \  }
 \\{}
 &&& &&&&\ddots
 \end {array}\right ]\,.
 \een
Similarly, we took  $k=3$ in the next continuation  of the series
of solutions $\Theta_k^{(QL)}$ pertaining to $H_k$. It is easy to
verify that the three-parametric quasi-local solution
$\Theta_3^{(QL)}$ of eq.~(\ref{tarov}) is still obtainable as a
diagonal matrix with the same elements $(1-a)\,(1-b)^2\,(1-c)^2$
in all the upper left corner and with the similar array of the
same elements $(1+a)\,(1+b)^2\,(1+c)^2$ in its lower right corner.
The remaining ``central" quadruplet of the ``anomalous" diagonal
elements is formed by the following four-dimensional diagonal
central submatrix of our doubly infinite matrix $\Theta_3^{(QL)}$,
 \ben
 \left [\begin {array}{rrccrr}%
&\multicolumn{2}{l}{\!\!\!(1-a)\,(1-b)^2\,(1-c^2)\ \ \ \ }&&&
 \\%
 &&\multicolumn{2}{r}{\ \ \ \ \ \ \ \ (1-a)\,(1-b^2)\,(1-c^2)}&&
 \\%
 &&&\multicolumn{2}{c}{ (1+a)\,(1-b^2)\,(1-c^2)\ \ \ \ \ \ \ \ \ }
 &
 \\%
 &&& &\multicolumn{2}{c}{\ \ \ \ \ \ \ \ \ \  (1+a)\,(1+b)^2\,(1-c^2) }
 \\%
 \end {array}\right ]\,.
 \een
The general pattern of extrapolation is now obvious. It would be
easy to write down and, via eq.~(\ref{tarov}), verify an immediate
extrapolation of the $k=1$, $k=2$ and $k=3$ matrices
$\Theta_k^{(QL)}$ to the higher subscripts $k$ whenever necessary.

There are all reasons why, in the context of scattering, the
specific diagonal metrics $\Theta_k^{(QL)}$ should be preferred in
comparison with all their non-diagonal and, hence, more nonlocal
alternatives. With such a new postulate we may now return to paper
\cite{Jones} once more. First of all our results reconfirm the
high plausibility of the hypothesis that even the violation of the
Hermiticity which is strictly localized in space should be
expected to influence, manifestly, even the asymptotics of the
wave functions. The explicit analysis of our schematic models
indicates that even our minimally nonlocal metric operators
remain, strictly speaking,  different from the most common Dirac's
delta-function metric $\Theta^{(Dirac)}(x,y) = \delta(x-y)$ at all
distances.

This being said, we found it quite fortunate that at the
sufficiently large distances, i.e., for $|x| \gg 1$ and/or $|y|
\gg 1$, the difference between $\Theta^{(Dirac)}(x,y)$ and
$\Theta_k^{(QL)}(x,y)$ degenerated, in all of our models, to the
mere introduction of a nontrivial multiplication factor,
 \be
 \Theta^{(QL)}(x,y) =const({\rm sign}\, x)\cdot
 \Theta^{(Dirac)}(x,y)\,,
 \ \ \ \ \ \ \ \ |x| \gg 1\,,\ \ \ \ |y|\gg 1\,.
 \label{prope}
 \ee
In another formulation, our explicit constructions very strongly
support the {\em affirmative} answer to the question of the
existence of a ``spatially localized non-Hermiticity". A formal
key to such an answer is that in a schematic model we constructed
certain new and very specific, ``quasi-local" metrics
$\Theta^{(QL)}$ with the property (\ref{prope}).

The {\em strict} validity of this proportionality rule {at almost
all the coordinates} $x$ and $y$ may be admitted to be an artifact
resulting from our specific tridiagonal-matrix choice of our
``toy" Hamiltonians $H_k$. Still, the validity of such a rule at
all the sufficiently large coordinates may be expected to survive
transition to a larger family of models and, perhaps, also to some
slightly less friendly generalized quasi-linear forms of
$\Theta^{(QL)}$, say, with a band-matrix structure. This would
still allow us to conjecture that with the metrics
$\Theta=\Theta^{(QL)}$ the internal consistency of the models of
scattering (and, in particular, of their asymptotic boundary
conditions) would not be violated after the extension of the
present theory towards many less schematic and reasonably
non-Hermitian models of dynamics.

One of the instrumental versions of our conjecture will have the
form of the requirement $D'=0$ in  boundary conditions so that
eq.~(\ref{sctbc}) $\equiv $ eq.~(\ref{scatbc}). Then, the choice
of a {\em unique} metric $\Theta^{(QL)}$ characterized by its
minimalized nonlocality should be perceived as strongly
recommended in the conceptually consistent fundamental scattering
theory using non-Hermitian Hamiltonians.

Marginally, the latter requirement can be supported also by the
remark that ``it has been known for some time that ... for the
potential ${\rm i}x^3$ and the infinite ${\cal PT}-$symmetric
square well ... the particle is confined ... so that the range of
the non-locality is limited. Scattering potentials highlight this
feature [of nonlocality] to its full extent because the wave
functions ... do not have compact support." \cite{Jones}. This
means that the ``traditional" choices of $\Theta \neq
\Theta^{(QL)}$ can still offer a fully consistent model of the
physical reality for bound states. After all, we already noticed
that many models with $\Theta \neq \Theta^{(QL)}$ found
applications in nuclear physics \cite{Geyer} and in field theory
\cite{Carl}. Other constructions of $\Theta \neq \Theta^{(QL)}$
appeared also in the coupled-channel problems \cite{cc} or in the
Klein-Gordon-type models \cite{KG} etc.

\section{Summary \label{summ} }

In the differential-equation model of scattering studied in
ref.~\cite{Jones} the behavior of the physical ``in" and ``out"
states was strongly non-local so that, for example, the outgoing
waves contained a non-negligible ``incoming" component. In this
context we showed here that such a paradox is not inevitable for
non-Hermitian systems with real spectra. A set of counter-examples
has been described here in which the local non-Hermiticites
carried by the Hamiltonian implied just the necessity of the
replacement of the usual scalar product (\ref{usuall}) by the {\em
local, re-scaling} change of the measure,
 \be
 \br \psi|\psi'\kt = \int_{I\!\!R^2}\,\psi^*(x)\,
 \Theta(x)\,\psi'(x)\,dx\,.
 \label{newhnn}
 \ee
This enabled us to address several conceptual difficulties as
encountered in ref.~\cite{Jones} where the description of
scattering caused by several short-range non-Hermitian sample
potentials $V(x)$ has been presented. In this context we
discovered and described a family of non-Hermitian short-range
Hamiltonians $H_1$, $H_2$, $\ldots$ for which the description of
the scattering looks almost as easy and natural as in the standard
Hermitian regime.

Our selection of short-range interaction models proved technically
much simpler than expected. We revealed several amazingly close
parallels with their continuous delta-function analogues. We were
able to bring new arguments, first of all, thanks to certain
``unreasonable efficiency" of our non-perturbative  method. In
this framework, our main mathematical result is that the
hermitizing metrics $\Theta=\Theta_k$ which we attached to $H_1$,
$H_2$ and $H_3$ and, by an easy extrapolation, to any $H_k$ are
all represented by the (infinite-dimensional) {\em diagonal}
matrices. We believe that this is not just a friendly feature of
our specific models but rather a generic property of the metrics
since one has a lot of freedom of their modification in general.

In the latter spirit, our present main recommendation is that for
all the realistic non-Hermitian models of scattering one should
still {\em try to insist} on the requirement that the physical
metric $\Theta$ is {\em not too non-local}. In our present text we
succeeded in supporting the latter recommendation by a series of
the explicit illustrations of its {\em feasibility}. One of
reasons was that we choose the discretization of the real line of
coordinates as our principal methodical tool.

The first hints offering a background for such a decision were
already found and formulated in \cite{Jednap}. The present results
can briefly be characterized as a successful transition from the
bound-state models (or, formally, from the finite $N-$point
lattices of ref.~\cite{I}) to the scattering scenario (or,
formally, to the limit $N \to \infty$), complemented by the
replacement of the simplest possible one-parametric model of
ref.~\cite{Jednap} by the whole set of dynamically nontrivial
localized Hamiltonians $H_k$ containing, in principle, an
arbitrary finite number $k$ of coupling constants.

A formal benefit of our choice of the models appeared to lie in
their two-faced solvability.  Its first face was rather technical
and concerned an easiness of construction of the reflection and
transition coefficients. Certain massive cancellations in the
linear algebraic matching conditions made the final formulae
unbelievably compact. The second friendly face of the solvability
emerged during our systematic construction of the metrics
$\Theta$. An easiness of the guesswork encountered during
extrapolations $k \to k+1$ is worth mentioning since it proved
helpful and saved computer time.

{\it A priori} we couldn't have hoped in the amazing diagonality
of our solutions of eq.~(\ref{tarov}) or in their asymptotically
constant form or in a ``user-friendliness" of the transition from
the trivial model $H_k$ with $k=1$ to virtually all of its $k>1$
descendants. We firmly believe that at least some of these
properties will also be encountered in some other, similar but
less schematic models of the dynamics.

Needless to add that many emerging questions remain open. Some of
them (like, typically, the numerical efficiency of the
discretizations  and an analysis of the practical rate of their
convergence) have been skipped intentionally. The omission of some
other points was only made with regret, mainly because of their
lack of any immediate relevance for physics. For example, a
marginal but interesting benefit of the discretization with a
fixed gap $h>0$ could have been seen in the emergence of
parallelism between the discrete-lattice formulae and their
continuous-limit counterparts. Besides such a direct possible
correspondence between $H_k$s and point interactions, another
correspondence (viz., to the truncated, finite lattices) has also
been omitted as too mathematical, in spite of its potential
relevance for the verification of the reality of the spectra.

We are sure that even within the domain of physics we did not list
all the open questions. {\it Pars pro toto}, let us sample, in the
conclusion, the possible relevance of the present models with the
extremely simple metrics in the context of the path-integral
formulation of Quantum Theory where an extremely interesting
discussion just appears in print \cite{path}, concerning the
questions of the role of the explicit form of the metric $\Theta$
in the partition functions $Z[J]$ and in certain related formulae
in thermodynamics and/or quantum field theory.

 \vspace{5mm}

\section*{Acknowledgement}

Numerous inspiring discussions of the subject with Hugh Jones are
gratefully appreciated. Work supported by the M\v{S}MT ``Doppler
Institute" project Nr. LC06002,  by the Institutional Research
Plan AV0Z10480505 and by the GA\v{C}R grant Nr. 202/07/1307.


\newpage
 \begin{thebibliography}{99}

\bibitem{Jones}
H. F. Jones, Phys. Rev. D 76, 125003 (2007).

\bibitem{BB}
C. M. Bender and S. Boettcher,
Phys. Rev. Lett. 80, 5243 (1998).

\bibitem{BG}
V. Buslaev and V. Grecchi, J. Phys. A: Math. Gen. 26, 5541 (1993).

%

\bibitem{Geyer}
F. G. Scholtz, H. B. Geyer and F. J. W. Hahne, Ann. Phys. (NY)
213, 74 (1992).


\bibitem{Carl}
C. M. Bender, Rep. Prog. Phys. 70, 947 (2007).


\bibitem{Ahmed}
Z. Ahmed, Phys. Lett. A 324, 152 (2004);

Z. Ahmed, C. M. Bender and M. V. Berry,
 J. Phys. A: Math. Gen. 38, L627 (2005);


M. Znojil,
J. Phys. A: Math. Gen. 39, 13325 (2006);

F. Cannata, J.-P. Dedonder and A. Ventura, Ann. Phys. 322, 397
(2007).
%
%
%
%

\bibitem{Jonesdva}
H. F. Jones, Interface between Hermitian and non-Hermitian
Hamiltonians in a model calculation, arXiv:0805.1656 [hep-th] 12
May 2008.

\bibitem{Jednap}
M. Znojil, submitted.

\bibitem{delta}
A. Mostafazadeh, J. Phys. A: Math. Gen. 39, 13495 (2006).


\bibitem{footnotea}
This operator is denoted here by the symbol $\Theta$ because this
is just a Greek-letter translation of its oldest denotation by $T$
in \cite{Geyer}. We prefer this symbol not only to $T$ (which
might associate the time reversal) and to equivalent symbols
$\eta$ and $\exp \,Q$ \cite{Jones} but also to Bender's most
popular but rather specific product ${\cal CP}$ \cite{Carl} and to
Mostafazadeh's subscripted $\eta_+$ \cite{Ali}. All of these
symbols carry precisely the same meaning but none of them seems to
be generally accepted yet.

\bibitem{SIGMA}
M. Znojil,
%
SIGMA 4,  001 (2008).
%

\bibitem{cc}
M. Znojil, J. Phys. A: Math. Gen. 39, 441 (2006);

M. Znojil, Phys. Lett. A 353, 463 (2006).

\bibitem{Ali}
B. Bagchi, C. Quesne and M. Znojil,
Mod. Phys. Lett. A 16, 2047  (2001);

A. Mostafazadeh, J. Math. Phys. 43, 205  (2002);

H. Langer and Ch. Tretter, Czech. J. Phys. 54, 1113 (2004);

M. Znojil and H. B. Geyer, Phys. Lett. B 640, 52 (2006).


\bibitem{I}
M. Znojil, Phys. Lett. A 223, 411 (1996);

F. M. Fern\'andez, R. Guardiola R, J. Ros  and M. Znojil, J. Phys.
A: Math. Gen. 31, 10105 (1998);

M. Znojil, J. Phys. A: Math. Gen. 39, 10247 (2006).

\bibitem{jaotom}
M. Znojil, Phys. Lett. A 203, 1 (1995).
%


%

\bibitem{dva}
M. Znojil, J. Phys. A: Math. Theor. 40, 4863 and 13131 (2007).



\bibitem{Jonespr}
H. F. Jones, private communication.

\bibitem{KG}
A. Mostafazadeh, Class. Quant. Grav. 20, 155 (2003) and Ann. Phys.
309, 1 (2004);

M. Znojil, H. B\'{\i}la and V. Jakubsk\'{y}, Czech. J. Phys. 54,
1143 (2004);

F. Kleefeld, Czech. J. Phys. 56, 999 (2006).

\bibitem{path}
V. Jakubsk\'{y}, Mod. Phys. Lett. A 22, 1075 (2007);

H. F. Jones and R. J. Rivers, Phys. Rev. D 75, 025023 (2007);

A. Mostafazadeh, Phys. Rev. D 76, 067701 (2007);

M. Znojil, Math. Reviews, extended abstract Nr. 2358121
(submitted).


\end{thebibliography}
\end{document}